\documentclass[useAMS,usenatbib]{mn2e}

\usepackage{graphicx}
\usepackage{txfonts}
\usepackage{natbib}
\usepackage{longtable}
\usepackage{amssymb}
\newcommand{\kms}{km\,s$^{-1}$}

\title[Detection of three triple stars in open clusters]{Dynamical detection of three triple stellar systems in open clusters}
\author[J. F. Gonz\'alez,  M. E. Veramendi, and C. R. Cowley]
{J. F. Gonz\'alez$^{1}$\thanks{E-mail:fgonzalez@icate-conicet.gob.ar}, 
M. E. Veramendi$^{1}$, and 
C. R. Cowley$^{2}$\\
$^{1}$Instituto de Ciencias Astron\'omicas, de la Tierra y del Espacio, Casilla de Correo 49, 5400 San Juan, Argentina\\
$^{2}$Department of Astronomy, University of Michigan, Ann Arbor, MI 48109-1042, USA}
\begin{document}

\date{Accepted 1988 December 15. Received 1988 December 14; in original form 1988 October 11}

\pagerange{\pageref{firstpage}--\pageref{lastpage}} \pubyear{2002}

\maketitle

\label{firstpage}

\begin{abstract}
We present a kinematic analysis of three  triple stellar systems belonging to
two open clusters: CPD$-$60$^\circ$961 and HD\,66137 in NGC\,2516, and HD\,315031 in NGC\,6530.
All three systems are hierarchical triples with a close binary bound to a third body in a 
wider orbit, whose presence is detected through  velocity variations of the 
close binary barycentre.
Orbital parameters are derived from radial velocity curves. 
Absolute parameters for all stars are estimated assuming cluster membership.
Some dynamical and evolutionary aspects of these systems are discussed, particularly the
possible influence of Kozai cycles. 
The two systems of NGC\,2516 have similar orbital configurations with inner periods of 11.23~d and 
8.70~d and outer periods of 9.79~yr and 9.24~yr. 
The young system HD\,315031 in the cluster NGC~6530 has an inner binary with a period of 1.37~d 
and a very eccentric ($e$=0.85) outer orbit with a period of 483~d. 
We report also radial velocity measurements of the components of the visual binary  CPD$-$60$^\circ$944 in NGC\,2516.
Including results from previous works, this cluster would harbor 5 hierarchical triples.
Possible dynamical evolutionary scenarios are discussed.
Long-term radial velocity monitoring is highlighted as strategy for the detection of subsystems with intermediate separations, which are hard to cover with normal spectroscopic studies or visual techniques.

\end{abstract}


\section{Introduction}

The frequency and the observed properties of multiple systems constitute key
evidence for models of stellar formation and evolution. 
Presently, our statistical knowledge of early-type high-order multiple systems 
is insufficient for a useful test of theoretical models.
In fact, in most of the catalogued multiple systems, the hierarchical configuration,
the physical link between the companions, and even the number of stellar components
has not been reliably established. 
Indeed, in our spectroscopic study of stellar components in 30
early-type multiples \citep{2014A&A..xxx..xxxV}, we found that a significant fraction 
of the systems have a number of components that is different from those previously known.  

Early-type multiple stars are also crucial for the understanding of B and A-type chemically peculiar stars,
whose physics is not completely understood.
On the one hand, peculiar stars that occur in multiple systems and clusters may be 
assigned an age, which makes it possible to approach the time development of
chemical peculiarities \citep{2014A&A...561A.147B}. 
The study of the companions gives eventually the opportunity  
to know the original composition of the peculiar star. On the other hand,
statistically some types of chemical peculiarities appear predominantly in binary or multiple
stars. This is the case of HgMn stars for example, whose multiplicity frequency might be as 
high as 90\% \citep{2010A&A...522A..85S}.
Finally, there is observational evidence suggesting that stellar companions might influence 
the generation of surface regions with abnormal composition
\citep{2006MNRAS.371.1953H,2012A&A...547A..90H,2011A&A...529A.160M}.

Open clusters provide large samples of coeval stellar objects. However, 
the studies of multiplicity in open clusters are surprisingly scattered in the literature 
\citep{2013ARA&A..51..269D}. Searches for spectroscopic and visual binaries have been carried 
out only in the nearest associations (Taurus-Aurigae, the Hyades, Praesepe, and the Pleiades), 
for which frequencies and some properties of multiple systems have been derived. Currently, 
hydrodynamic simulations of cluster formation \citep{2012MNRAS.419.3115B} predict fractions 
of multiplicity increasing with the stellar mass, with values in agreement with the 
observational results for solar- or low-mass stars, including very-low-mass objects.

According to \citet{2011MNRAS.410.2370L}, dynamical encounters involving triple systems should 
be common in open clusters and they would significantly contribute to the population of 
blue-stragglers. \citet{2013MNRAS.432.2474L} confirmed these predictions considering the fractions 
of binaries and triples empirically determined in Taurus-Aurigae, the Hyades, Praesepe, 
and the Pleiades. These authors concluded that in such clusters  encounters involving 
triple systems are as frequent as, or even more, than encounters involving single stars and 
binaries. These results suggest that models of open cluster evolution should include high-order 
multiples in their initial population and allow  these systems to evolve dynamically through 
the cluster evolution and to influence the dynamical evolution of the latter through energy 
exchange \citep{2013MNRAS.432.2474L}. From the observational point of view, these results highlight 
the need for a more detailed knowledge of the binary and especially triple population in 
clusters \citep{2011MNRAS.410.2370L}.
 
 The detection and exhaustive characterization of hierarchical multiples represent
an observational challenge, since different instrumental techniques have to be
combined to cover all the range of separation between the components of subsystems,
which span several orders of magnitudes from a few solar radii to thousands of astronomical units.
Spectroscopy is the main tool for studying binary subsystems with periods
shorter than tens or a few hundreds of days, while
astrometric techniques and high-resolution imaging are used for the widest systems.
Particularly tough is the intermediate range of separations, with periods on
the order of a few thousands of days (a$\sim$10 AU). 
In the present paper we report the discovery of triple systems through the detection
of long-period variations in the centre-of-mass velocity of spectroscopic binaries,
demonstrating the usefulness of long-term spectroscopic monitoring for the study of
this intermediate range of separation.

In the long-term spectroscopic survey of stars in open clusters carried out 
at ICATE \citep{2004RMxAC..21..141L} several new spectroscopic binaries have been discovered.
Eventually, some of the systems showed run-to-run variations of the barycentric 
velocity and from then on were monitored until the whole cycle of those slow variations was covered. 

The cluster NGC\,2516 is particularly interesting for its content of binary and peculiar stars.
Early spectroscopic works of this cluster reported about ten chemically peculiar 
late-B and early-A type stars \citep{1969AJ.....74..813A,1972A&A....21..373D,1976ApJ...205..807H}
and nine single-lined spectroscopic binaries \citep{1972ApJ...172..355A,1982A&AS...49..497G}.
However, only a few of these objects have been studied subsequently and good quality orbits
have been published for only two of them: the eclipsing binary V392~Car \citep{2001A&A...374..204D}
and HD\,65949 \citep{2010MNRAS.405.1271C}. 
On the other hand, \citet{2000AJ....119.2296G} reported three new double-lined spectroscopic 
binaries (SB2s) and one radial velocity variable (CPD$-$60$^\circ$944A).
The stars studied in the present paper, HD\,66137 and CPD$-$60$^\circ$961 
\citep[stars 19 and 2 in the cluster numbering by][]{1955ApJ...121..628C}, are
two of these three SB2s.
The third one is HD66066A, a short period binary studied by \citet{2003A&A...404..365G}, who
mentioned that would be a triple star, since it has a visual companion to which it could be dynamically 
bound \citep{1989A&AS...78...25D}.
On the other hand, no spectroscopic study of stars HD\,66137 and CPD$-$60$^\circ$961  
has been published since they were reported as binaries, although 
preliminary results of the present investigation, were presented by \citet{2010ASPC..435..107V}.
In addition to these two objects, in the present paper we present a few observations 
of CPD$-$60$^\circ$944A, the star reported as variable in \citet{2000AJ....119.2296G}, 
confirming its binary nature. Since this star has (at least) one visual companion \citep{1989A&AS...78...25D}
this would be also a multiple star.

HD\,315031 is an early-B type star member of the young open cluster NGC\,6530.
It was studied by \citet{2003A&A...404..365G}, who measured the radial velocity
of both components using two-dimensional cross-correlations 
\citep{1994ApJ...420..806Z}, and determined the spectroscopic orbit.
Already in that first study, the measurements from different runs were found to be inconsistent
with a constant binary centre-of-mass velocity, 
which was interpreted as due to the presence of a third body.
However, the  available data were not sufficient to secure the period of the outer orbit.

In Sect.~\ref{sec:obs} and \ref{sec:param} we describe the orbital analysis and derive
physical properties of the systems studied. In Sect~\ref{sec:disc} we discuss some dynamical
and evolutionary aspects. Finally, our main conclusions are summarised in Sect.~\ref{sec:conc}.

\section[]{Observations and orbital analysis}\label{sec:obs}

In the present paper we analyse spectroscopic observations taken over 15 years
with the 2.15 m telescope and the REOSC echelle spectrograph at Complejo Astron\'omico El Leoncito (CASLEO), San Juan, Argentina. 
The spectra cover the wavelength range 3700--5900~\AA~with a resolving power R=13\,300. 
The spectra were reduced by using standard procedures with the NOAO/IRAF package. 
In the case of  HD\,315031 our analysis includes remeasurement of most of the spectra of \citet{2003A&A...404..365G}.

Our spectra show, in principle, two sets of spectral lines belonging to the companions
of the close binary system. 
We will identify these stars with letters A and B, being A the most massive of this pair.
The third companion in the outer orbit will be called star C, regardless of its mass.   
We applied the spectral separation method by \citet{2006A&A...448..283G} to
measure radial velocities and reconstruct the spectra of both visible stellar components.
For the cross-correlations used by this method in the RV calculation,
standard templates taken from the Pollux database \citep{2010A&A...516A..13P} were used.
Table\,\ref{tab:rvs} (available electronically) lists all measured radial velocities for
the three spectroscopic binaries.
Columns 3 and 4 are orbital phases for the inner and outer orbits, calculated from the time of periastron passage.
In the case of the inner orbit of HD\,315031, which is circular, the phase origin is at the primary conjunction.
Columns 5 to 8 list the measured radial velocities and their errors. 
Columns 9 and 10 are the calculated velocities
for the two close-binary companions with respect to its barycentre, while the last column lists
the calculated radial velocity for the barycentre of the binary, $V_\rmn{o}$.

\begin{table*}
 \centering
 \begin{minipage}{140mm}
  \caption{Radial velocities. This is a sample of the full table.}\label{tab:rvs}
  \begin{tabular}{@{}ccccrrrrrrr@{}}
  \hline
   \multicolumn{1}{c}{Object} & \multicolumn{1}{c}{HJD}  & 
   \multicolumn{1}{c}{$\phi$} & \multicolumn{1}{c}{$\phi_\rmn{o}$} & 
   \multicolumn{1}{c}{$V_\rmn{A}$} & \multicolumn{1}{c}{err$_\rmn{A}$}& 
   \multicolumn{1}{c}{$V_\rmn{B}$} & \multicolumn{1}{c}{err$_\rmn{B}$} & 
   \multicolumn{1}{c}{$(V_\rmn{A}-V_\rmn{o})_\rmn{cal}$}&   \multicolumn{1}{c}{$(V_\rmn{B}-V_\rmn{o})_\rmn{cal}$}&  \multicolumn{1}{c}{$V_\rmn{o}$}\\ 
   &  & & & \kms &\kms & \kms& \kms & 
   \multicolumn{1}{c}{\kms} &   \multicolumn{1}{c}{\kms}&  \multicolumn{1}{c}{\kms}\\ \hline
CPD$-$60$^\circ$961    &2451587.7233  &0.9437  &0.4843  &16.4  &0.6  &39.6  &0.6  &-9.7  &14.0  &26.8 \\
CPD$-$60$^\circ$961    &2451590.6421  &0.2035  &0.4851  &47.1  &0.6  &-5.3  &0.7  &21.4  &-30.8  &26.8 \\
CPD$-$60$^\circ$961    &2451939.7566  &0.2838  &0.5819  &46.9  &0.5  &-1.2  &0.5  &19.4  &-28.0  &26.6 \\
CPD$-$60$^\circ$961    &2452270.7007  &0.7465  &0.6736  &6.3  &0.8  &56.0  &1.1  &-20.6  &29.7  &26.2 \\
CPD$-$60$^\circ$961    &2452271.7615  &0.8409  &0.6739  &5.1  &0.8  &54.6  &0.9  &-20.3  &29.3  &26.1 \\

... &&&&&&&&&\\
\hline
\end{tabular}
\end{minipage}
\end{table*}

Radial velocity data were fitted with  a double Keplerian orbit.
In this scheme, the two visible components are assumed to move in a two mass-points orbit
whose centre-of-mass follows in turn a Keplerian orbit.
In the orbital phase calculation, observation times are corrected by the
light-time effect due to the variation of the distance between the close binary and the
observer. 
To find the best solution our orbit fitting program seeks in the 
parameter space the minimum of the quantity $\chi^2$.  Consequently data points
are weighted according to their formal measurement errors.
The parameter errors are calculated considering the parameter variation for an 
increment $\Delta \chi^2$=1 while all remaining parameters are allowed to vary to minimize $\chi^2$, 
taking into account in this way correlations between parameters.
In this manner, we fitted simultaneously 12 orbital parameters:
the six elements of the inner orbit
(orbital period $P$, time of periastron passage $T_\pi$, 
radial velocity amplitudes $K_\mathrm{A}$ and $K_\mathrm{B}$,
argument of periastron $\omega$, and eccentricity $e$) and 
six parameters for the outer orbit that the binary barycentre 
 describes around the centre-of-mass of the whole triple system
(systemic velocity $\gamma_o$, period $P_\mathrm{o}$, time of periastron passage $T_{\pi\mathrm{,o}}$, 
radial velocity amplitude $K_\mathrm{o}$,
argument of periastron $\omega_\mathrm{o}$, and eccentricity $e_\mathrm{o}$).
Table\,\ref{tab:orb} lists the orbital parameters for the three systems
including, in addition to the mentioned orbital elements, the time of
primary conjunction $T_\mathrm{I}$, projected orbital semiaxis $a\sin i$,
and minimum masses $M\sin^3 i$. 
Note that the subindex ``o'' does not correspond to the orbit of star C but to
the observed movement of the barycentre of the spectroscopic binary. 
In particular, the semiaxis of the outer relative orbit 
  is $a_\rmn{out}=a_\rmn{o}\cdot(M_\rmn{A}+M_\rmn{B}+M_\rmn{C})/M_\rmn{C}$.
For notation simplicity in the parameters of the inner binary concerning both components
($P$, $e$, $\omega$, $q$, etc.) we use no subindex.

\begin{figure}
   \centering
   \includegraphics[height=0.95\linewidth,angle=-90]{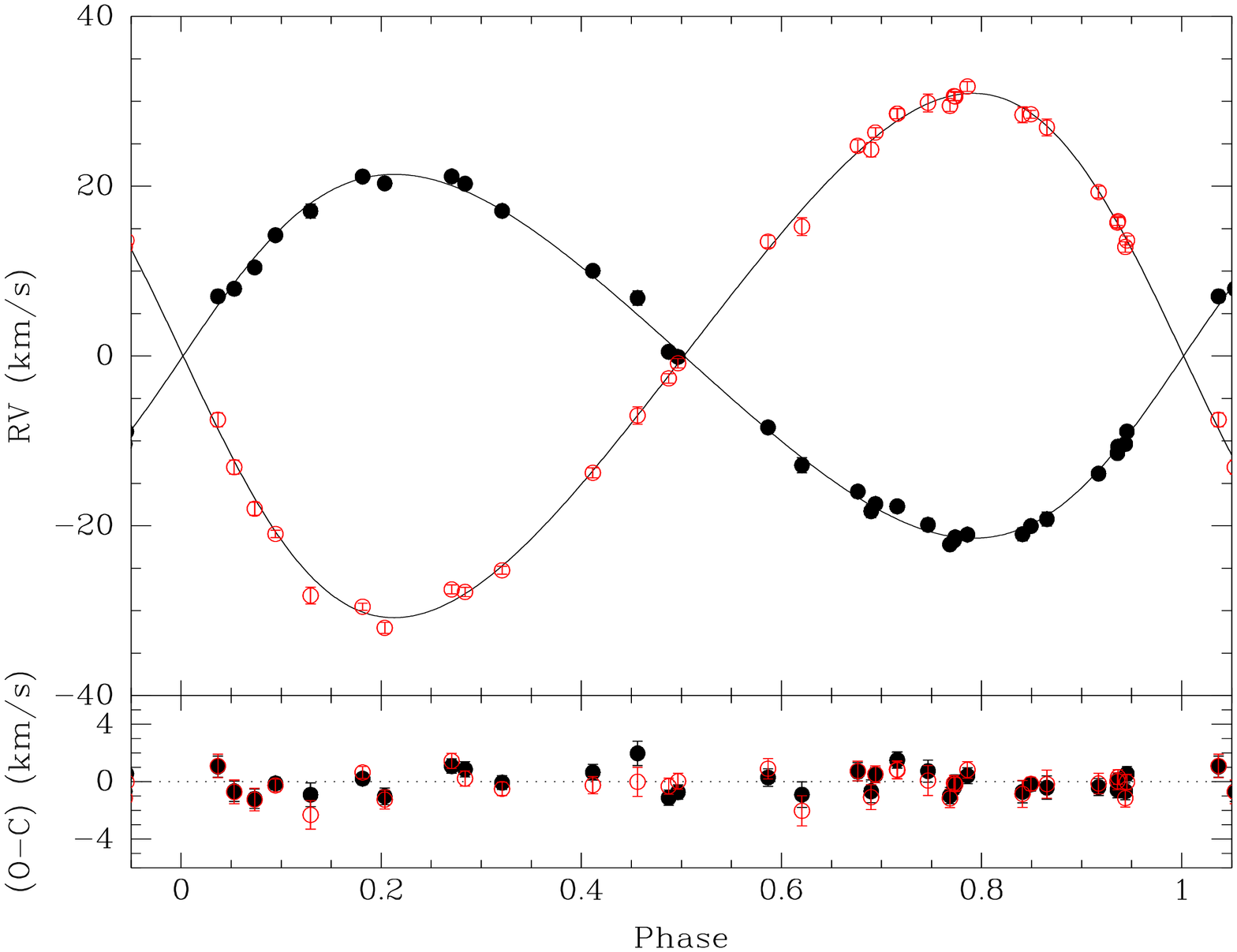} %
   \includegraphics[bb= 2 2 450 737, width=4.5cm,height=0.95\linewidth,angle=-90]{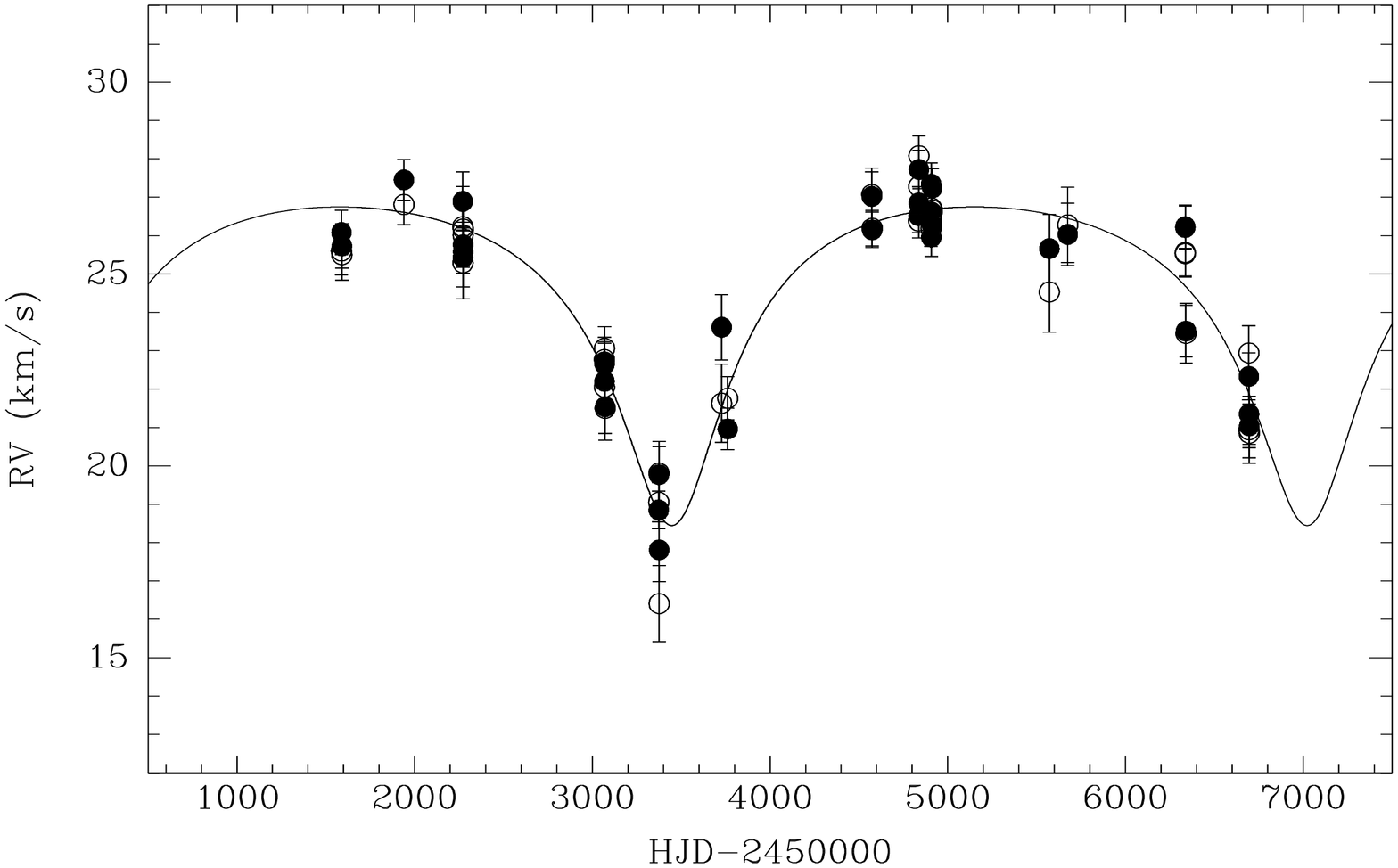}
   \caption{Radial velocity curves of  CPD$-$60$^\circ$961. 
   The upper panel shows the radial velocity curves of the inner binary over
   the 11.23 day period.
   At the bottom of the same panel residuals observed-minus-calculated are plotted.
   The  lower panel shows the longer time variation of the centre-of-mass velocity.
   Filled (open) circles correspond to the primary (secondary) star.}
         \label{rv02}
         
   \end{figure}

\begin{table*}
\caption{Orbital parameters. Parameters with subindex "o" refers to the outer orbit 
described by the barycentre of the spectroscopic binary. The last
three rows are the number of observations and the RMS of the residuals.}
\label{tab:orb}
\centering
\begin{tabular}{lcr@{ $\pm$ }lr@{ $\pm$ }lr@{ $\pm$ }l} 
\hline \hline
Parameter	&	Units	&	 \multicolumn{2}{c}{CPD$-$60$^\circ$961}	&			  \multicolumn{2}{c}{HD\,66137}	&			 \multicolumn{2}{c}{HD\,315031}\\ \hline
$P$			&	d			&	11.23259	&	0.00013	&	8.70355	&	0.00004	&	1.377436	&	0.000003	\\
$T_\mathrm{I}$	&	d			&	2\,454\,705.43	&	0.04	&	2\,454\,446.884	&	0.014	&	2\,454\,484.623	&	0.005	\\
$T_\pi$		&	d			&	2\,454\,688.55	&	0.13	&	2454448.142	&	0.017	&	\multicolumn{2}{c}{-}	\\
$K_\rmn{A}$ 	&	km\,s$^{-1}$		&	21.4	&	0.3	&	64.4	&	0.7	&	66.2	&	1.1	\\
$K_\rmn{B}$		&	km\,s$^{-1}$		&	30.9	&	0.3	&	64.7	&	0.6	&	89.0	&	1.6	\\
$\omega$			&				&	4.70	&	0.07	&	3.236	&	0.014	&	\multicolumn{2}{c}{-}\\
$e$			&				&	0.124	&	0.009	&	0.380	&	0.005	&	 \multicolumn{2}{c}{0.00}	\\
$q$	&       &	0.694	&	0.013	&	0.995	&	0.013	&	0.744	&	0.018	\\
$a_\rmn{A}\sin i$	&	R$_\odot$   &	4.72	&	0.07	&	10.24	&	0.10	&	1.80	&	0.03	\\
$a_\rmn{B}\sin i$	&	R$_\odot$   &	6.80	&	0.07	&	10.29	&	0.10	&	2.42	&	0.04	\\
$a\sin i$	&	R$_\odot$	&	11.52	&	0.10	&	20.53	&	0.14	&	4.22	&	0.05	\\
$M_\rmn{A}\sin^3 i$	&	M$_\odot$	&	0.096	&	0.003	&	0.769	&	0.016	&	0.306	&	0.012	\\
$M_\rmn{B}\sin^3 i$	&	M$_\odot$	&	0.067	&	0.002	&	0.765	&	0.016	&	0.228	&	0.008	\\
$\gamma_\rmn{o}$		&	km\,s$^{-1}$		&	24.54	&	0.19	&	24.64	&	0.24	&	6.8	&	0.8	\\
$P_\rmn{o}$		&	d			&	3575	&	65	&	3376	&	39	&	482.9	&	0.5	\\
$T_\rmn{I,o}$	&	d			&	2\,456\,385	&	55	&	2\,456\,138	&	46	&	2\,454\,513	&	13	\\
$T_\rmn{\pi,o}$	&	d		&	2\,460\,611	&	148	&	2\,454\,865	&	125	&	2\,455\,339	&	2	\\
$K_\rmn{o}$		&	km\,s$^{-1}$		&	4.2	&	0.4	&	10.1	&	0.4	&	37.6	&	1.6	\\
$\omega_\rmn{o}$     &				&	3.2	&	0.2	&	5.4	&	0.3	&	5.15	&	0.13	\\
$e_\rmn{o}$		&				&	0.47	&	0.06	&	0.17	&	0.03	&	0.85	&	0.03	\\
$a_\rmn{o}\sin i_\rmn{o}$	&	R$_\odot$	&	258	&	21	&	660	&	24	&	190	&	10	\\
$n$	&	&	 \multicolumn{2}{c}{34}	&	 \multicolumn{2}{c}{36}	&	 \multicolumn{2}{c}{71}	\\
$\sigma_\rmn{A}$	&	km\,s$^{-1}$	&	 \multicolumn{2}{c}{0.58}&	 \multicolumn{2}{c}{1.17}&	 \multicolumn{2}{c}{4.4}\\
$\sigma_\rmn{B}$	&	km\,s$^{-1}$	&	 \multicolumn{2}{c}{0.61}&	 \multicolumn{2}{c}{1.23}&	 \multicolumn{2}{c}{7.7}\\
\hline
\end{tabular} 
\end{table*}

For both triples of NGC\,2516, CPD$-$60$^\circ$961 and HD\,66137, 
cluster membership is confirmed kinematically by the agreement between barycentric velocity
of the two triples and the cluster velocity: 22.0 \kms \citep{2000AJ....119.2296G}, 23.08 \kms \citep{2008A&A...485..303M}.

\begin{figure}
   \centering
   \includegraphics[angle=-90,width=0.95\linewidth]{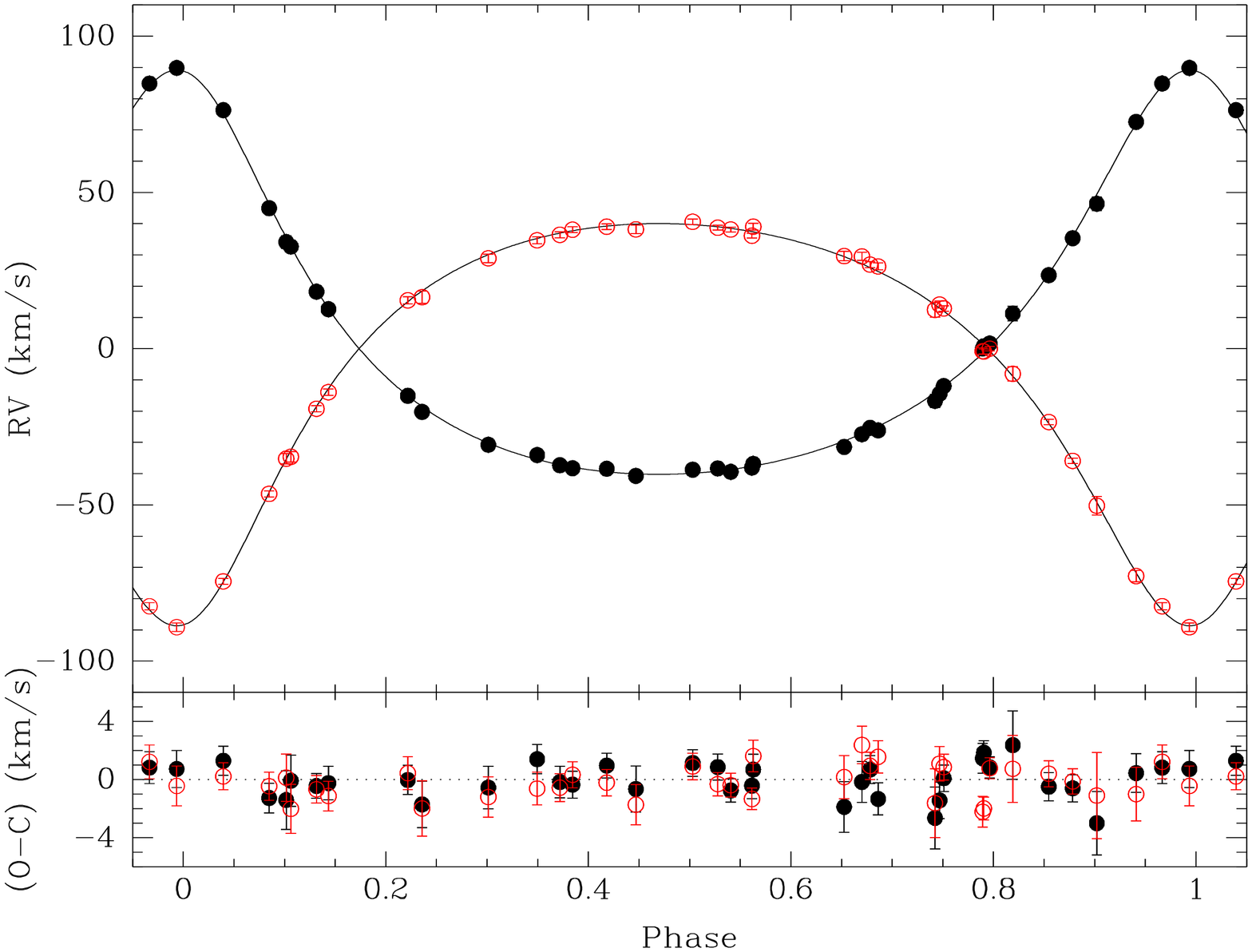}
   \includegraphics[bb= 2 2 450 737, width=4.5cm,height=0.95\linewidth,angle=-90]{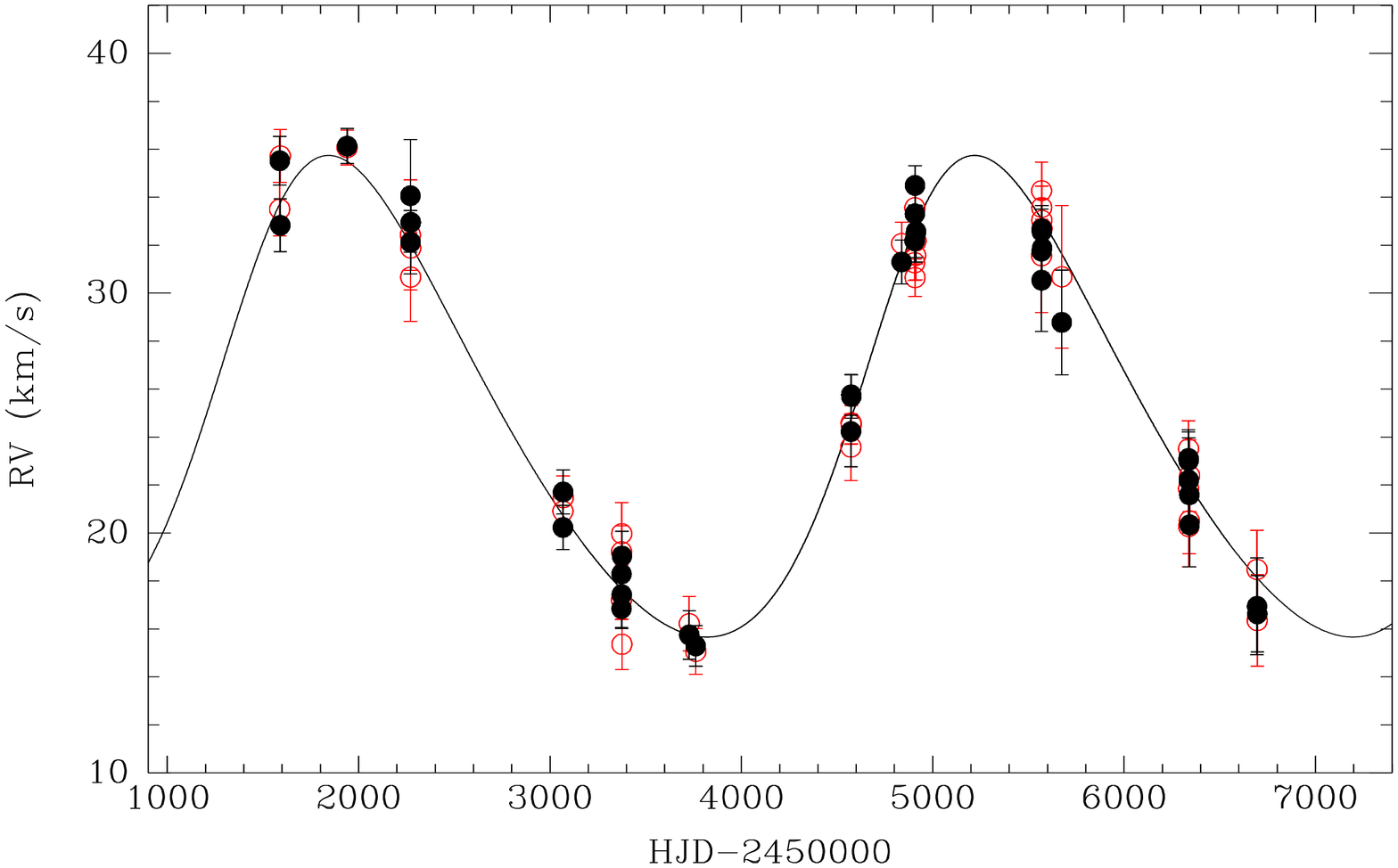}
      \caption{Radial velocity curves of  HD\,66137. 
   The upper panel shows the radial velocity curves of the inner binary.
   At the bottom of the same panel the residuals observed-minus-calculated are plotted.
   The  lower panel shows the time variation of the centre-of-mass velocity.
   Filled (open) circles corresponds to the primary (secondary) star.}
         \label{rv19}
   \end{figure}

\begin{figure}
   \centering
   \includegraphics[height=0.95\linewidth,angle=-90]{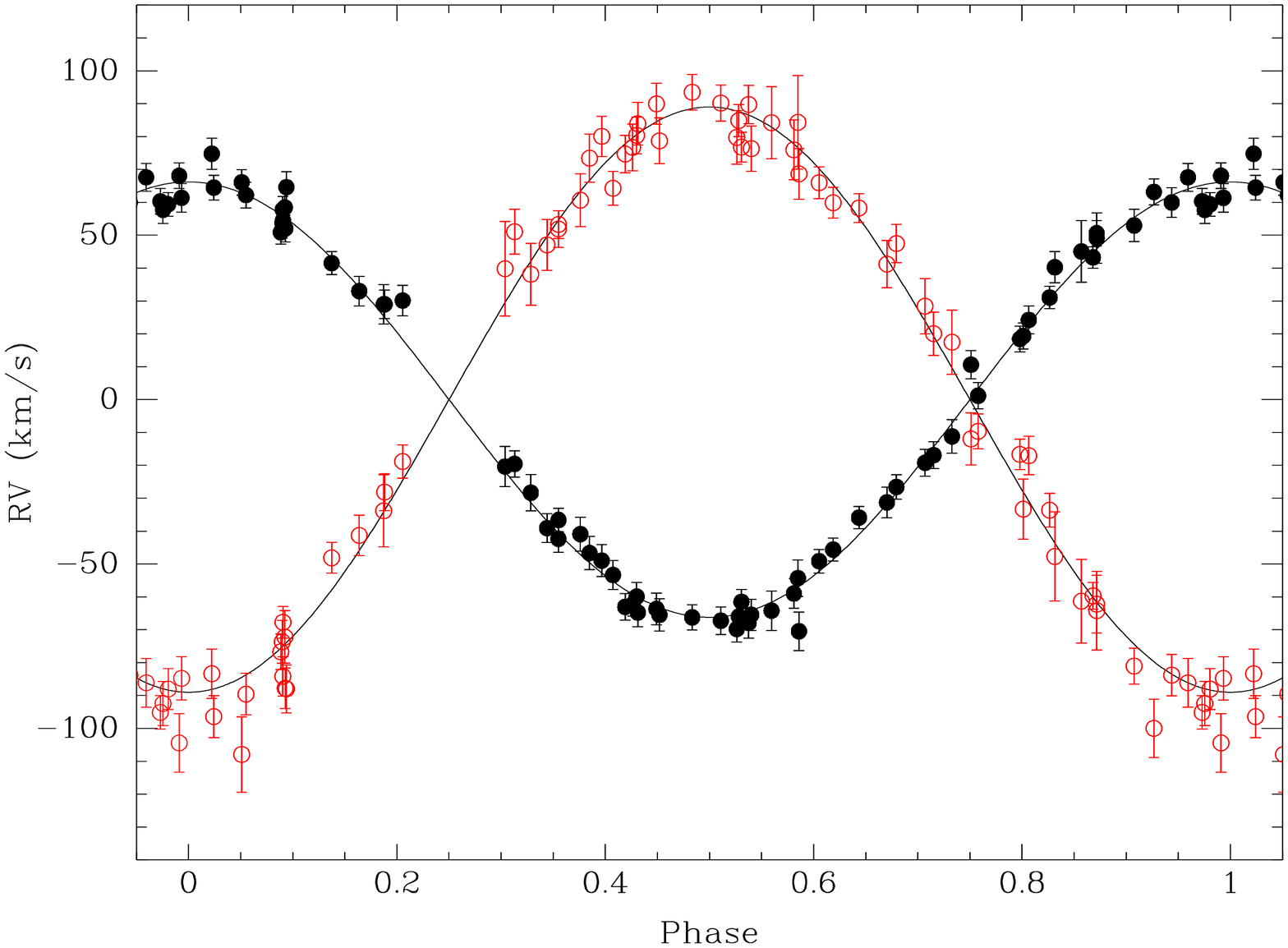}
   \includegraphics[bb= 2 2 450 737,width=5cm,height=0.95\linewidth,angle=-90]{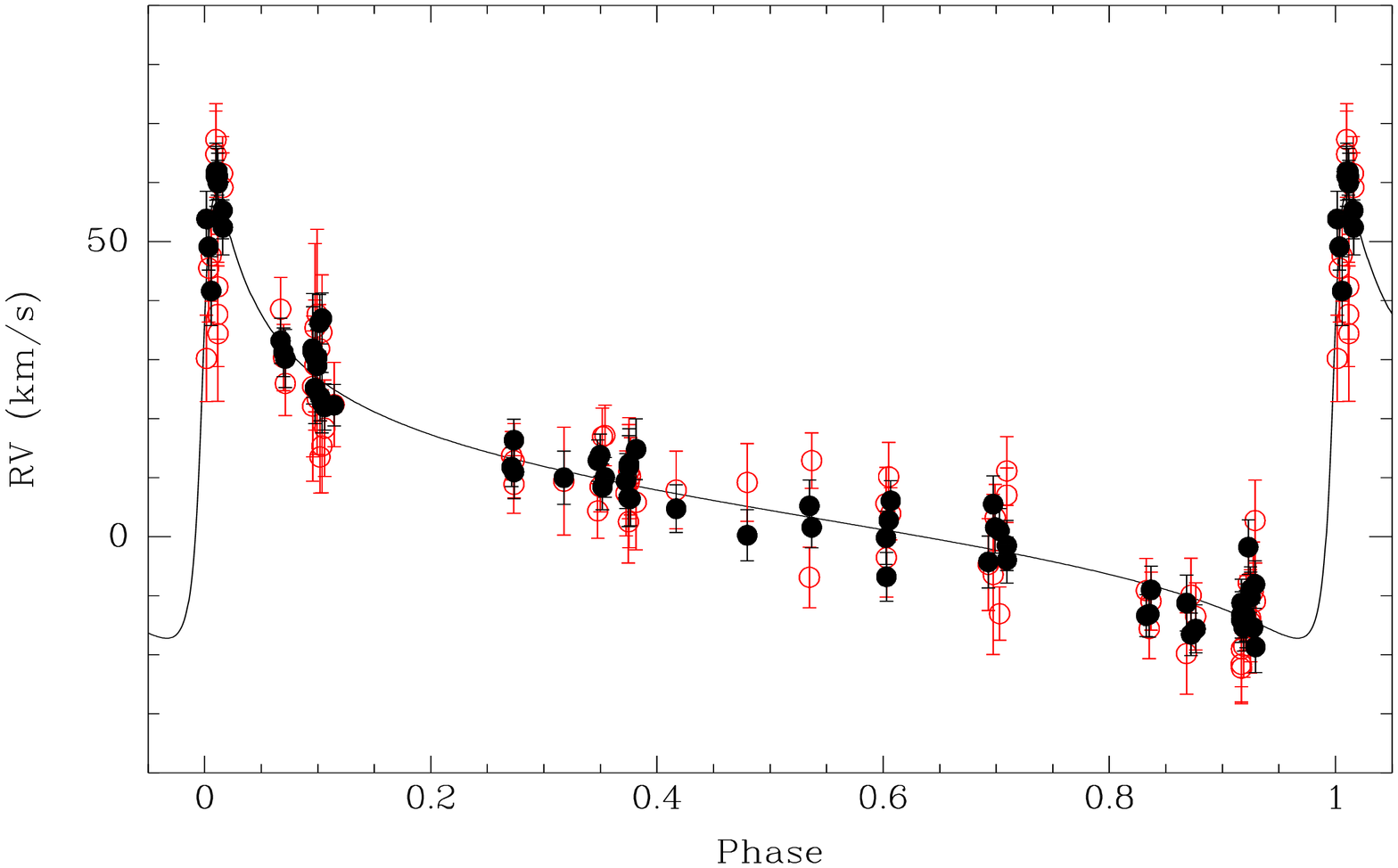}
      \caption{Radial velocity curves of  HD\,315031. 
   The upper panel shows the radial velocity curves of the inner binary.
   At the bottom of the same panel the residuals observed-minus-calculated are plotted.
   The  lower panel shows the time variation of the centre-of-mass velocity.
   Filled (open) circles corresponds to the primary (secondary) star.}
         \label{rv73}
   \end{figure}

Radial velocity curves are shown in Figures \ref{rv02}, \ref{rv19}, and \ref{rv73}. 
In these figures upper panels show the velocity of the spectroscopic binary companions
with respect to the barycentric velocity of the binary, while the lower panels show the velocity variations of the binary barycentre.

\section{Physical parameters}\label{sec:param}

Cluster membership allows us to know the age of these multiple systems and to
estimate absolute stellar parameters from their position in the cluster color-magnitude diagram.
Nevertheless, the third body might contribute significantly 
to the integrated light of the system, making these calculations less reliable.
Therefore, we adopted primarily estimated masses from the spectral types.
Using line ratios measured in the separated spectra, we determined spectral types for 
the six studied stars.
We then derived temperatures using the  \citet{Schmidt-Kaler} calibration and 
estimated masses and other parameters interpolating in the isochrone corresponding to the
cluster age in the stellar model grids of
the Geneva group \citep{2012A&A...537A.146E,2012A&A...541A..41M,2013A&A...553A..24G}.
Uncertainties in these parameters were estimated from the 
range of spectral types  considered compatible with the observed spectrum.

We used the spectroscopic parameters of the outer orbit and the estimated masses of the
visible companions to derive a lower limit to the mass of the third body.
Combining the expression for the radial velocity semiamplitude of the wide orbit:
$$
K_\rmn{o} = \frac{2\pi a_\rmn{o}}{P_\rmn{o}}\frac{\sin i_\rmn{o}}{\sqrt{1-e_\rmn{o}^ 2}}\leq \frac{2\pi a_\rmn{o}}{P_\rmn{o}\sqrt{1-e_\rmn{o}^ 2}}
$$
with the Kepler equation for the outer orbit we obtain:
\begin{equation}\label{eq1}
\frac{M_\rmn{C}}{(M_\rmn{A}+M_\rmn{B}+M_\rmn{C})^{2/3}}=\frac{K_\rmn{o} P_\rmn{o}^{1/3} \sqrt{1-e_\rmn{o}^2}}{(2\pi G)^{1/3}\sin i_\rmn{o}}\geq \frac{K_\rmn{o} P_\rmn{o}^{1/3} \sqrt{1-e_\rmn{o}^2}}{(2\pi G)^{1/3}}.
\end{equation}

On the other hand, when it was possible we estimated an upper limit to $M_\rmn{C}$ 
considering the intensity that its spectral lines would have in the spectrum. 
This upper limit to $M_\rmn{C}$ was used to obtain a lower limit to the outer orbit inclination $i_\rmn{o}$
using eq.~\ref{eq1}.
Obtaining at least rough estimates for the masses of all three stars, allow us to get,
in combination with the spectroscopic parameters, information about the mutual inclination
between the inner and the outer orbits.
In fact, the mutual inclination between both orbital planes ($i_\rmn{m}$) is given by:
$$
\cos(i_\rmn{m}) = \cos(i) \cos(i_\rmn{o})+\sin(i) \sin(i_\rmn{o}) \cos(\Delta\Omega),
$$
where $\Delta\Omega$ is the angle between the lines of nodes of the two orbits.
Therefore, 
$$
 \cos(i_\rmn{m}) \leq \cos(i-i_\rmn{o}),
$$
and finally, taking into account that both $i_\rmn{m}$ and $\left|i-i_\rmn{o}\right|$ are
in the interval (0, $\pi$), we obtain
$$
 i_\rmn{m} \geq \left|i-i_\rmn{o}\right|.
$$

In short, from estimates of stellar masses it is possible to obtain a lower limit for the  
inclination between inner and outer orbits, which is an important parameter for the dynamics of the triple.
In Table~\ref{tab:i} we summarise these results, which are discussed for each system in the following sections.

\begin{table}
\centering
 \begin{minipage}{80mm}
 \caption{Estimated absolute masses and inferred orbital inclinations. }
\label{tab:i}
\centering
\begin{tabular}{lcr@{ $\pm$ }lr@{ $\pm$ }lr@{ $\pm$ }l} 
\hline \hline
Parameter	&	Units	&	 \multicolumn{2}{c}{CPD$-$60$^\circ$961}	&			  \multicolumn{2}{c}{HD\,66137}	&			 \multicolumn{2}{c}{HD\,315031}\\ \hline
$T_\mathrm{eff}\mathrm{(A)}$&	K		&	10\,500	&	1000	&	10\,100	&	900	&	28\,000	&	2000	\\
$T_\mathrm{eff}(B)$		&	K			&	8450	&	250	&	10\,100	&	900	&	25\,400& 2000	\\
$M_\rmn{A}$ 	&	M$_\odot$		&	2.6	&	0.4	&	2.3	&	0.3	&	12.9	&	1.5	\\
$M_\rmn{B}$		&	M$_\odot$		&	1.8	&	0.2	&	2.3	&	0.3	&	9.6	&	1.1	\\
$i$			&		deg		&	19.5	&	1.0	&	43.5	&	2.6	&	16.7 & 0.6 \\
$M_\rmn{C}$\footnote{Lower limit from the spectroscopic mass-function.}		&	M$_\odot$	&	
    \multicolumn{2}{c}{$>0.8$}	&	 \multicolumn{2}{c}{$>2.6$}	&	\multicolumn{2}{c}{$>7.0$}	\\
$M_\rmn{C}$\footnote{Estimated from detection/non-detection of lines in the spectrum.}&	M$_\odot$	&
	\multicolumn{2}{c}{$<1.6$}	&	\multicolumn{2}{c}{$\sim 3.2\pm 0.5$}	&	 \multicolumn{2}{c}{$<10$}	\\
$i_\rmn{o}$				&		deg		& \multicolumn{2}{c}{$\ga 72$} & \multicolumn{2}{c}{$\sim 77\pm 7$}& \multicolumn{2}{c}{$\ga 75$}\\
$i_\rmn{m}$		&	deg		& \multicolumn{2}{c}{$\ga 52$}	&\multicolumn{2}{c}{$\ga 34$}&\multicolumn{2}{c}{$\ga 58$}\\
\hline
\end{tabular} 
\end{minipage}
\end{table}

\subsection{CPD$-$60$^\circ$961}\label{sec:star2}

For the components of this spectroscopic binary we derived spectral types B9\,V and A4\,V.
In our spectra both stars have sharp lines (full-with half-maximum of about 25 \kms) with
rotational velocity too low to be measured with our spectral resolution.
As already noted in the preliminary work by \citet{2010ASPC..435..107V}, star A shows a peculiar spectrum with a 
notable $\lambda$3984 Hg\,\textsc{ii} line and other lines typical of HgMn stars (strong Y\,\textsc{ii}, 
Sr\,\textsc{ii},  P\,\textsc{ii}, and Pt\,\textsc{ii}). 
However, strikingly, Mn\,\textsc{ii} lines are not present.
We analyzed 884 wavelength measurements by the method 
of wavelength coincidence statistics and  
the results shows very low probability for the presence of Mn.
P\,\textsc{ii} is clearly present but not Ga\,\textsc{ii}.
Highly significant results were obtained for Sr\,\textsc{ii}, Y\,\textsc{ii}, and Pt\,\textsc{ii}. 
Interestingly, the presence of several lanthanide spectra is very likely, particularly
lines of Nd\,\textsc{iii} and Pr\,\textsc{iii}.
All these characteristics, which altogether are not typical of any group of peculiar stars, resemble
the spectrum of the unique star HD\,65949 studied by \citet{2006A&A...455L..21C,2010MNRAS.405.1271C},
which is also a spectroscopic binary of the same cluster NGC\,2516.

From the spectral types we estimated temperatures $T_\rmn{eff}(\rmn{A})$=10\,500$\pm$1000\,K and
$T_\rmn{eff}(\rmn{B})$=8\,450$\pm$250\,K.
We then interpolated in the Geneva model grid to find stellar models that are consistent with
the temperature estimates, the spectroscopic mass-ratio, and the cluster age.
We adopted for the cluster $\log (\rmn{age})$=8.0--8.3, which is a conservative range consistent with 
the published values: 8.04 \citep{1989A&AS...78...25D}, 8.15 \citep{1993A&AS...98..477M}, 
8.2$\pm$0.1 \citep{2002AJ....123..290S}.
We obtained masses $M_\rmn{A}=2.6\pm 0.4$~M$_\odot$ and $M_\rmn{B}=1.8\pm 0.2$~M$_\odot$, and
visual absolute magnitudes $M_{V}(\rmn{A})=0.9\pm 0.5$ mag and $M_V(\rmn{B})=2.3\pm 0.2$ mag.
The comparison of these masses with the spectroscopic parameters gives
a low orbital inclination ($i$=19\fdg 5$\pm$1\fdg 0) and for $M_\rmn{C}$ a lower limit of 0.8~M$_\odot$.
An upper limit can be established considering that there is no trace of its lines in the spectrum. 
Assuming this third body is a main-sequence star, 
we estimate that its relative flux is at least 30\% lower than the flux of the secondary star,
even if its lines were rotationally broadened.
Therefore, $M_\rmn{C}$ is expected to be less than about 1.6~M$_\odot$.

Assuming an absolute magnitude corresponding to a star of 0.8--1.6~M$_\odot$ for component C,
we obtained an integrated absolute magnitude for the system of about $M_\rmn{V}$(ABC)=+0.6$\pm$0.4 mag,
which corresponds to an apparent distance modulus $V-M_\rmn{V}=8.2\pm 0.4$ mag.
This value is consistent with the photometric values determined for the cluster by 
\citet[][8.54 mag]{1989A&AS...78...25D}, 
\citet[][8.12 mag]{2000AJ....120..333S}, \citet[][8.30 mag]{2002ApJ...576..950T}, and
\citet[][8.19]{1994A&AT....4..153L}, as well as the one derived from the Hipparcos parallaxes 
\citep[$V_\rmn{o}-M_\rmn{V}=7.70\pm 0.16$ mag, ][]{1999A&A...345..471R}.

Approximate values for the  orbital inclination of both the inner and outer orbits 
can be derived if the masses of the three stars are estimated.
For the inner orbit we obtained $i\approx$19\fdg 5. 
For the outer orbit, assuming an upper limit of 1.6 M$_\odot$ for star C, we
calculated a lower limit  $i_\rmn{o} \ga 72^\circ$.  
Therefore, the mutual inclination of the inner and outer orbits is $i_\rmn{m} \ga  52^\circ$.

Dynamically the system contains a close binary with two stars about 2.6 M$_\odot$ and 1.8 M$_\odot$ in an  
orbit with a semiaxis of about 35 R$_\odot$ and moderate eccentricity. 
This close pair is bound to a third star, less massive,  
orbiting in an eccentric orbit with a semiaxis of about 8~AU. 
The triple is markedly hierarchical with  a semiaxis ratio of $\sim$50.

\subsection{HD\,66137}

The spectral morphology of components A and B are very similar to each other:
both have spectral type B9-A0\,V and low rotational velocity. 
Their temperatures would be approximately 9700$\pm$400 K, which
in the cluster isochrone corresponds to a mass of 2.3$\pm$0.3~M$_\odot$
and an absolute visual magnitude of 1.35$\pm$0.45 mag.
The  lower limit for the mass of star C is high.
Using eq.~\ref{eq1} we deduced that 
star C  contains at least 35--37\% of the total mass of the triple, being probably
the most massive star of the three.
Likewise, the intensity of spectral lines of the two visible components suggests that the flux 
contribution of star C is considerable.
Indeed, the equivalent width of all spectral lines in stars A and B are smaller
than expected for stars of the same spectral type by a factor 0.3--0.4.
In absence of a third light, this factor would be 0.5 for both stars. 
Therefore, all three stars seem to have comparable brightness.

\begin{figure}
   \centering
   \includegraphics[width=0.95\linewidth]{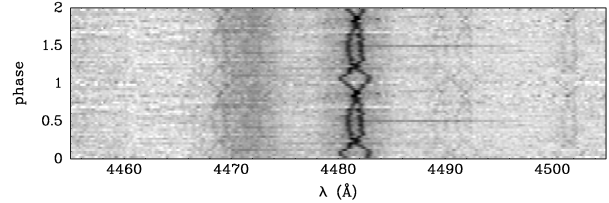} 
      \caption{Stacked spectra of HD\,66137 ordered by orbital phase.
      Shadows at $\lambda$4471 and  $\lambda$4481 are broad spectral
      lines belonging to the fast rotating star C. 
      Several sharp lines of the close binary are visible:  the strong line Mg\,\textsc{ii} 4481 and
      weak lines of Ti\,\textsc{ii} (4468 and 4501) and Fe\,\textsc{ii} (4489 and 4491).}
         \label{stack}
   \end{figure}
   
The lines of star C are in fact present in the spectrum, but pass unnoticed 
 due to its high rotation. In order to make its spectrum more visible, we 
built the gray-scale image shown in Fig.~\ref{stack}, in which all observed spectra (excluding
a few low-S/N spectra) are stacked ordered by orbital phase.
Besides the sharp lines of the spectroscopic binary, which nicely trace
its orbital motion, 
broad spectral lines are visible as shadows at the position of 
lines He\,\textsc{i} $\lambda$4471 and Mg\,\textsc{ii} $\lambda$4481. 
A few other similar broad lines can be distinguished in the spectrum, including  He\,\textsc{i} lines
at $\lambda\lambda$4089, 4144, 4388,\ He\,\textsc{i}-Fe\,\textsc{ii} blends at $\lambda$4922-24 and
$\lambda$5016-18, and the Si\,\textsc{ii} doublet $\lambda$4128-30.
Since  He\,\textsc{i} $\lambda$4471 and Mg\,\textsc{ii} $\lambda$4481 have similar intensity,
we estimate a spectral type B7-B8 for star C, somewhat earlier than its companions.
Its rotational velocity is on the order of 200-250 \kms.
If these broad lines belong indeed to the same object that is responsible for the barycentric 
movement of the close binary, its radial velocity is expected to vary with a semiamplitude of the order
of 10 \kms.  These variations, however, are very hard to measure with such shallow and broad lines.

The integrated magnitude of the triple allow us to make an estimate of the mass
of star C. From the apparent magnitude of the triple \citep[$V=7.85$,][]{1989A&AS...78...25D},
the apparent distance modulus, 
and the estimated magnitudes of the components
of the inner binary, we obtain for star C an absolute magnitude 
$M_\rmn{V}(\rmn{C})\approx$+0.2 mag, which corresponds to a mass $M_\rmn{C}\sim$3.2 M$_\odot$.
From the estimated mass of star C we derived  $i_\rmn{o} \approx 77^\circ\pm 7^\circ$ and
 a lower limit $i_\rmn{m} \ga 34\degr$ for the mutual inclination of the orbits, 
although considering the uncertainties in $i_\rmn{o}$ and $i$ this lower limit might be
 as low as 24\degr.

In short, this triple is structured in two hierarchical levels. 
The outer binary subsystem has a period of 9.2 yr and a semiaxis of about 8--9 AU.
The less massive component of this subsystem is a 3 M$_\odot$ star, while the primary 
is a close binary of 8.7 d period with twin components of about 2.3 M$_\odot$.

\subsection{HD\,315031}\label{sec:73}

We determined for the visible components of this system's spectral types B0.5\,IV-V and B1\,V,
which correspond to temperatures $T_\rmn{eff,A}=28\,000\pm 2000$~K and  
$T_\rmn{eff,B}=25\,400\pm 2000$~K, according to the \citet{Schmidt-Kaler} calibration.
 The low values of the spectroscopic minimum masses in comparison with the masses corresponding
 to the spectral types, indicate a low orbital inclination.

The eccentricity of the orbit is indistinguishable 
from zero, as is expected as a consequence of tidal friction
for a binary with a period as short as this (P=1.377 d). 
Hence, in the final calculation of the orbital parameters we fixed
the eccentricity at zero.
Furthermore, both stellar components are expected to rotate synchronously with 
the orbital motion, since the timescale for synchronization is shorter
than for circularization.
Under this hypothesis, the projected radii can be calculated from the projected
rotational velocities.
We determined $v \sin i$ by applying the method by \citet{2011A&A...531A.143D} on
 the reconstructed spectra of stars A and B.
 We obtained $v_\rmn{A} \sin i = 48\pm 4$ \kms and $v_\rmn{B} \sin i = 42\pm 4$, from which
 we calculated $R_\rmn{A} \sin i= 1.31 \pm 0.11\:\rmn{R}_\odot$
 and  $R_\rmn{B} \sin i= 1.14 \pm 0.12\:\rmn{R}_\odot$.
Even though the orbital inclination is in principle unknown, the projected radius
can be combined with the minimum mass obtained from the orbital analysis to
get the mean stellar densities:
\begin{equation}\label{eq2}
\left(\frac{\rho}{\rho_\odot}\right)_j= 0.01343 \ \left[\frac{K_\mathrm{A}+K_\mathrm{B}}{(v\sin i)_j}\right]^3\cdot \frac{q^{j-1}}{P^2\:(1+q)} 
\end{equation}
where $j$=1 ($j$=2) for star A (B), the period is in days, 
and the numerical constant is $(1\:\mathrm{R}_\odot/1\:\mathrm{AU})^3\cdot (1\:\mathrm{yr}/1\:\mathrm{d})^2$.

\begin{figure}
   \centering
   \includegraphics[angle=-90,width=0.95\linewidth]{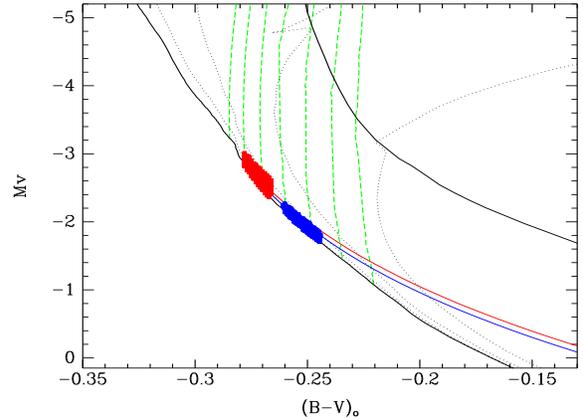} 
      \caption{Position of the stellar components of HD\,315031 in the Color magnitude diagram.
      Black heavy lines are the zero-age and terminal-age main-sequences; dotted lines
      are the isochrones $\log \tau$=6.0, 6.5, 7.0, and 7.5; dashed lines are isotherms
      from 20\,000~K (right) to 32\,000~K (left); thin continuous lines are
      curves corresponding to densities $\rho = 0.137\: \rho_\odot$ and $\rho = 0.152\: \rho_\odot$.
      The gray (red and blue in the color version) areas mark the stellar 
      models compatible with the observed temperatures and densities.}
         \label{fig:hr}
   \end{figure}

Using this equation we obtained: 
$\rho_\mathrm{A} = 0.137\pm 0.035\: \rho_\odot$ and $\rho_\mathrm{B} = 0.137\pm 0.035\: \rho_\odot$.
These values correspond to stars very close of the zero-age main-sequence.
Fig.~\ref{fig:hr} shows the position of both companions in the Color-Magnitude diagram.
The colored regions mark the stellar models that are consistent with all observational
information, essentially  spectroscopic mass-ratio, densities, and temperatures. 

In short, this binary is formed by two unevolved main-sequence stars of 
12.9$\pm$1.5 and 9.6$\pm$1.1~M$_\odot$ with radii of about 
4.6$\pm$0.4 and 3.9$\pm$0.3~R$_\odot$, corresponding to an age of about 1 Myr.
The inclination of the orbit is about 16\fdg 7.
The spectroscopic minimum mass for the third star is a fraction 0.24$\pm$0.01 of the
total mass. This value corresponds to 0.7-0.8 times the mass of star B.
Considering that the lines of star C are not clearly visible in the spectrum,
its mass should be close to this lower limit.
Even if it is a fast rotator, a conservative higher limit for its mass would be 9-10 M$_\odot$.

The integrated absolute visual magnitude derived from the estimated stellar parameters is
in agreement  with the cluster membership.
Although NGC\,6530 has been subject of several photometric studies,
the distance to this cluster is not well known. 
Published values for the distance modulus range from $V_0-M_\rmn{V}\approx 11.3$ to $V_0-M_\rmn{V}\approx 10.5$ \citep{2000AJ....120..333S,2006MNRAS.366..739A,2005A&A...430..941P}, while the $E(B-V)$ reddening value would be between 0.20 and 0.35 mag. The disagreement between
different authors or even different star samples within the same work, might be related 
to variable extinction, an abnormal extinction law, 
or the existence of several stellar groups at different distances \citep{2006MNRAS.366..739A}.
The location of this triple in the cluster color-magnitude diagram corresponds to
$M_\rmn{V}\approx -2.9$ for $V_0-M_\rmn{V}$=10.5, $E(B-V)$=0.2 and $M_\rmn{V}\approx -4.1$ for $V_0-M_\rmn{V}$=11.25, $E(B-V)$=0.35.
Our estimated absolute parameters correspond to an integrated absolute magnitude of the binary 
HD\,315031AB of $M_\rmn{V}= -3.16\pm0.25$. Assuming $M_\rmn{C}$ is in the range 7--10 M$_\odot$,
the total absolute magnitude of the triple would be $M_\rmn{V}\approx -3.4$, which is consistent
with the cluster distance.

The total mass A+B+C is of the order of 31~M$_\odot$, and therefore the outer orbital
semiaxis is about 800 $R_\odot$. However, since the orbit is very eccentric,
at periastron the third star is at only 120  $R_\odot$ from the close binary.
This distance is about 8 times larger than the separation of the companions
of the inner pair.   
The high inclination of the outer orbit and low inclination of the inner 
orbit ($i\approx 17^ \circ$) assure a high relative inclination of the two 
orbital planes  ($i_\rmn{m} \ga 58$\degr).

\section{Discussion}\label{sec:disc}

\subsection{Detection of long period spectroscopic subsystems}

We reported in the present paper the dynamical discovery of two triples
in the cluster NGC\,2516 with outer periods between 9 and 10 years.
Our time baseline was long enough to cover in both cases
about one and a half orbital cycles of the outer subsystems.
At the distance of the cluster (380 pc), the projected semiaxis of the outer subsystems 
in angular units is, in both triples, about 20--25 mas.  
This separation range is still out of the reach of visual techniques like speckle interferometry
or adaptive optic imaging, showing the difficulty of a complete multiplicity survey,
covering all separation ranges.
Even in close well-studied clusters like the Pleiades and Praesepe our knowledge of multiplicity
at intermediate separations (5--15 AU) is presently very poor \citep{2013MNRAS.432.2474L}.
Interestingly, the  semiaxis of the outer orbit for our two triples of NGC\,2516 are
in the middle of this separation range, showing the importance of long time-baseline 
radial velocity monitoring in multiplicity surveys.

Most long-period spectroscopic binaries in open clusters have been discovered
by studies focussed on late-type stars, like the systematic
survey of giant stars of \citet{1989A&A...219..125M} \citep[see also][]{2007A&A...473..829M}.
The small-amplitude velocity variations of wide binaries are easier to detect in late-type
stars where higher precision can be reached in radial velocity measurements.
By contrast, our knowledge of early-type wide binary systems is much poorer,
as is evident from the content of the Ninth Catalogue of 
Spectroscopic Binary Orbits \citep{2009yCat....102020P}.
Among the 254 binaries with periods above 3000 days in this catalog,
231 are late-type (F-M) dwarfs or giants, 
while only 14 binaries have A-type primaries and 9 have spectral types O or B.
Binaries and multiples with periods of several years are virtually impossible to discover in
fast rotating early-type stars. However, errors of 1 \kms can be obtained in O-B-A stars
of low $v\sin i$, and the monitoring such objects,
particularly in young clusters, would alleviate the lack of 
statistical information on massive long period spectroscopic binaries.

\subsection{Triple stars of NGC\,2516}

In this section we show, through the analysis of published and new information, that the
cluster NGC~2516 would contain five hierarchical triples among its early-type main-sequence stars.
The two triple systems detected dynamically in the present paper are hierarchical with similar 
orbital configuration: an outer subsystem with
a semiaxis of about 8 AU and an inner subsystem with $a\approx 30\:\rmn{R}_\odot$.
However, in the triple CPD$-$60$^\circ$961 the most distant star (component C) is the least massive
of the three, while in HD\,66137 it is the most massive.

Additionally, we present here spectroscopic material for one visual-spectroscopic triple.
The visual binary CPD$-$60$^\circ$944AB is very probably a physical pair \citep{1989A&AS...78...25D}, 
and its primary was reported as radial velocity variable by \citet{2000AJ....119.2296G}
on the basis of two radial velocity measurements. 
We present in Table\,\ref{tab:208-209} velocity measurements obtained in the last years for both
visual companions. This observations definitely confirm the binarity of the visual primary. 
The available observations, however,  are not sufficient to fit the orbit reliably. We found
two possible orbits with periods of 121.6 and 182.5 days and eccentricity $e\approx 0.4$ in both
cases. 
A notable fact of this triple is that both visible components are chemically peculiar.
CPD$-$60$^\circ$944A was reported as B8pSi by \citet{1976ApJ...205..807H}.
In our spectra Si\,\textsc{ii} is clearly enhanced, Cr\,\textsc{ii} appears weak, and Eu\,\textsc{ii} lines at 
$\lambda\lambda$ 4130, 4205, and 4436 \AA ~ are clearly visible.
Wavelength coincidence statistics showed  that several other rare earth elements  
are also present. With a wavelength tolerance of 0.06 \AA, 
highly significant results were obtained for ions Si\,\textsc{ii}, Ca\,\textsc{ii}, Ti\,\textsc{ii}, Fe\,\textsc{ii},
Pr\,\textsc{ii}, 
Nd\,\textsc{iii}, Eu\,\textsc{ii}, and 
Ho\,\textsc{ii}. 
At somewhat lower confidence level Cr\,\textsc{ii}, Nd\,\textsc{ii}, Dy\,\textsc{ii}, and Er\,\textsc{iii} would be also present. Even though significance levels change slightly using different tolerance values or line lists,
the presence of at least three lanthanide (Eu, Dy, Ho) is a robust result.
On the other hand, CPD$-$60$^\circ$944B, classified as B9.5IVp(Si) by \citet{1989A&AS...78...25D},  
is in fact a HgMn star, which exhibits strong lines of Hg\,\textsc{ii}, Mn\,\textsc{ii},
P\,\textsc{ii}, Ga\,\textsc{ii}, and Xe\,\textsc{ii}.
This system, therefore, would be one of the few known multiple system  formed by peculiar stars
of different type. 

\begin{table}
\caption{Radial velocities of visual pair CPD$-$60$^\circ$944AB. Uncertainties are
about 1.2 km\,s$^{-1}$ for component A and 0.9 km\,s$^{-1}$ for  component B.}\label{tab:208-209}
\centering
\begin{minipage}[t]{0.22\textwidth}
\centering
\begin{tabular}{cc} 
\multicolumn{2}{c}{CPD$-$60$^\circ$944A}\\
\hline 
HJD	&	RV  \\ 
	& km\,s$^{-1}$  \\
\hline
2\,451\,590.6287 &  22.7    \\
2\,451\,592.7282 &  24.3   \\
2\,451\,939.7416 &  18.7   \\
2\,452\,270.7174 &  \phantom{0}9.8     \\
2\,454\,905.5859 &   28.3   \\
2\,455\,572.6178 &  \phantom{0}9.8    \\
2\,456\,695.7500 &  22.8   \\ \hline
&\\
&\\
\end{tabular} 
\end{minipage}
\begin{minipage}[t]{0.22\textwidth}
\centering
\begin{tabular}{cc} 
\multicolumn{2}{c}{CPD$-$60$^\circ$944B}\\ \hline 
HJD	&	RV 	 \\
	& km\,s$^{-1}$ \\
\hline
2\,453\,761.7224 &  24.3    \\
2\,454\,835.6489 &  23.9    \\
2\,454\,905.5928 &  23.9    \\
2\,454\,907.5665 &  23.3     \\
2\,455\,566.6242 &  24.2    \\
2\,455\,572.6046 &  23.2     \\
2\,455\,674.5250 &  23.7     \\
2\,456\,695.7389 &  22.8     \\
2\,456\,696.6007 &  21.8    \\ \hline
\end{tabular} 
\end{minipage}
\end{table}

A fourth triple would be HD\,65949, a chemically-peculiar single-lined spectroscopic binary 
studied by \citet{2010MNRAS.405.1271C}, who detected a small variation in the centre-of-mass velocity,
which was interpreted as due to the presence of a third body. 
The chemical pattern of this star is not typical of any group
of peculiar stars, presenting several lines typical of HgMn stars, but lacking Mn lines.
Strikingly this morphology is very similar to the primary of  CPD$-$60$^\circ$961,
one of the triples analysed in this paper.

Finally, the short period spectroscopic binary HD66066A \citep{2003A&A...404..365G}
has a visual companion that could be dynamically bound \citep{1989A&AS...78...25D},
being therefore also a hierarchical triple.

\subsection{Dynamical evolution of the observed triples}

We will comment in this section about three aspects of the system dynamics: 
tidal effects in the close binary, orbital stability of the triple, 
and the possible occurrence of Kozai oscillations.
In close binaries, tidal interaction in combination with energy dissipation mechanisms 
(radiative damping in the case of early-type stars)
 tends to synchronize stellar rotation with orbital motion and circularize the orbit
\citep{1977A&A....57..383Z,1981A&A....99..126H}. 
In order to estimate the circularization timescales of the close pair in our triples
we used the binary evolution code developed by \citet{2002MNRAS.329..897H}\footnote{Publicly available at http://astronomy.swin.edu.au/$\sim$jhurley/.}.
Inner binaries in both triple systems of NGC\,2516 have periods long enough for  
circularization not to be reached until after the end of the main-sequence.
By contrast, HD\,315031 would have been circularized in about 1.0-1.5 Myr,
which is comparable with the cluster age.

According to \citet{2001MNRAS.321..398M}, a triple or higher-order star system is stable if the periastron distance of 
the third star (r$_\pi^\rmn{out}$) and the inner semiaxis ($a$) orbits obey the criterion:

$$
\frac{r_\pi^\rmn{out}}{a} > 2.8 \ \left[ (1+q_\rmn{out})\cdot
         \frac{1+e_\rmn{o}}{(1-e_\rmn{o})^{1/2}}\right]^{2/5}\cdot \left(1-0.3\frac{i}{\pi}\right),
$$
which as a function of the spectroscopic parameter $a_\rmn{o}$ can be written:
$$
\frac{a_\rmn{o}}{a} > 2.8 \ \frac{q_\rmn{out}}{(1+q_\rmn{out})^{3/5}}\cdot
         \frac{(1+e_\rmn{o})^{2/5}}{(1-e_\rmn{o})^{6/5}}\cdot \left(1-0.3\frac{i}{\pi}\right).
$$

For the three triples studied here this relation is satisfied.
In the two triples of NGC\,2516 the estimated semiaxis ratio is at least 10 times
larger than this minimum value for stability, while in the more eccentric system
HD\,315031 the observed value would be about twice the limit value.

In a hierarchical triple system, both the eccentricity of the inner binary and the mutual 
inclination execute periodic oscillations known as Kozai cycles 
\citep{1962AJ.....67..591K,1968AJ.....73..190H,2000ApJ...535..385F}.
The amplitude of the eccentricity variations are significant when
the relative inclination between the orbits is higher than $\arcsin(\sqrt{2/5})$=39\fdg 2,
being maximum when orbital planes are perpendicular to each other.
On the other hand, the eccentricity amplitude does not depend on either 
the mass of the third body or the outer semiaxis. 
The duration of the Kozai cycle, however, does depend on the outer period and the mass
of the distant star. 
In practice,  tidal effects between the companions of the inner binary, 
rotational deformation of the stars, or relativistic terms can detune the Kozai effect in systems
with large outer-to-inner semiaxis ratio. 
According to \citet{2009ApJ...703.1760M} the Kozai cycling is suppressed if
$P_\rmn{o}(\rmn{yr}) \gtrsim [P (\rmn{days}]^{1.4}$ .
In the three analysed systems the outer period is lower than this limit.

The period of Kozai cycles is given approximately by \citep{2000ApJ...535..385F,1979A&A....77..145M}:
\begin{equation}\label{eq3}
P_\rmn{e} \approx \beta \frac{P}{q_\rmn{out}} \left(\frac{a_\rmn{ou}}{a_\rmn{in}}\right)^3 (1-e^2_\rmn{o})^{3/2},
\end{equation}
where $\beta$ is a factor of order unity that depends on the initial values for 
mutual inclination,  eccentricity, and  argument of periastron of the inner orbit.
For the two triples of NGC\,2516 the periods of the eccentricity oscillations, 
which depend mainly  on orbital periods
$P$ and $P_\rmn{o}$, are similar to each other: $\sim$8--14 $\times 10^ 3$yr for
CPD$-$60$^\circ$961 and $\sim$8--10 $\times 10^ 3$yr for HD\,66137.
These values are much shorter than the timescales for tidal effects or nuclear evolution.
The short time-scale for the variation of orbital elements and the high mutual inclination
($i_\rmn{m}\ga 52^\circ$, see Sect.\,\ref{sec:star2}) 
assure that the Kozai mechanism  would be working efficiently in CPD$-$60$^\circ$961. 
The maximum eccentricity, which is a function of orbital inclination \citep{1999CeMDA..75..125K},
would be $e_\rmn{max}\ga 0.60$.
In the case of HD\,66137 the lower limit for the mutual inclination ($\ga 34\degr$)
 is slightly lower than the critical value, so the occurrence of Kozai
cycles in this system is not certain, although probable.
In fact, in both triples, mutual inclinations above 80\degr--85\degr are still compatible
with the spectroscopic parameters. If high-eccentricity configurations take place
periodically, the inner orbit could have been (or is being) shrunk by tidal interactions.
However, if during the eccentricity cycles the maximum values are not very high
($e_\rmn{max}\la 0.7$), then strong binary interactions (tidal circularization, mass-transfer)
would not take place before the end of the main-sequence stage.

If the inner orbit is not dynamically modified, in both triples the primary
star will overflow its Roche lobe during giant branch ascent before the core He-burning,
giving place to case B mass-transfer.

HD\,315031 has a very eccentric outer subsystem ($e_\rmn{o}$=0.85) and a very close inner subsystem ($P$=1.38\,d).
The angular momentum exchange between the inner binary and a third body in hierarchical 
triple systems has been proposed to play a key role in formation 
 of short-periods binaries \citep{1997AstL...23..727T}.
Statistics of binary and triples among solar-type stars supports this scenario \citep{2006A&A...450..681T}.
Even though the mutual inclination of the inner and outer orbits is high, we consider unlikely
that the inner binary has reduced its size
through the combination of Kozai cycles and tidal friction.
The reason is that the high eccentricity of the outer orbit (which remains constant in Kozai cycles), 
leaves little room for the size or the original inner orbit. 
For example, for an inner orbit with  $P\sim 6$~d the system would not be stable.  
A possible explanation is that the present configuration is  not the result of the
isolated evolution of the triple system, but the outcome of the dynamical decay of a higher order multiple system.
In fact, dynamical simulations of small-N clusters produce frequently triples with
relatively small period-ratio and high outer eccentricity 
\citep{2002A&A...384.1030S}, as is the case of HD\,315031.

This system is expected to experience mass-transfer during the main-sequence (case A mass-transfer).
In fact, the primary Roche lobe radius is currently $\sim$6~R$_\odot$, while the 
terminal-age main-sequence radius for a star with the primary mass is close to 12 R$_\odot$. 
According to Hurley's evolutionary code, the secondary star will become a blue straggler star
at about 14 Myr.  

\section{Summary and Conclusions}\label{sec:conc}

The long-term spectroscopic monitoring of three double-lined spectroscopic binaries
members of two open clusters, led to the discovery of the triple nature of these systems.
All three systems are hierarchical with a close pair and a third object in a wide orbit.
The ratio of the semiaxis of the outer and inner orbits are larger than 50, while the
period ratio is larger than 350.

Besides the two triples of NGC\,2516 analysed in detail in the present paper
(CPD$-$60$^\circ$961 and HD\,66137),
spectroscopic data for the visual-spectroscopic triple CPD$-$60$^\circ$944 of the same cluster
are reported. 
NGC\,2516 harbors five known hierarchical  triples with inner binaries in the spectroscopic
separation range, with periods between 1.7 and a few hundred days. Three of these systems
contains at least one chemically peculiar star. 

The triple HD\,66137 (and probably also CPD$-$60$^\circ$961) in NGC\,2516 might be
experiencing Kozai cycles with inner eccentricity oscillating in timescales of $\sim 10^4$yr. 
However, the orbits of the inner binaries have not been circularized yet.
Without a precise knowledge of the mutual inclination between the inner and outer orbit,
it is not clear if significant tidal effects are taking place at epochs of high inner eccentricity.
If the eccentricity remains lower than $\sim$0.7, strong tidal interaction is not expected
until the end of the main sequence.
Even without orbit shrinking, the inner binaries of both systems would experience 
case-B mass transfer, giving origin eventually to exotic objects.

On the other hand,  HD\,315031 contains a short-period massive binary and
a third star orbiting in a very eccentric orbit.
Due to the proximity of the third star at periastron the inner binary cannot have evolved 
from a significantly wider orbit. We speculate therefore that the present configuration 
is the result of a dynamical decay of a non-hierarchical multiple.
The inner binary is expected to suffer  mass transfer during the main-sequence stage
and the secondary star will become a cluster blue straggler star.

\section*{Acknowledgments}
This work was partially supported by a grant from FONCyT-UNSJ PICTO-2009-0125.
\bibliographystyle{mn2e}    
\bibliography{bibliomn}

\label{lastpage}

\end{document}